\documentclass[12pt,a4paper, preprint, preprintnumbers]{article} %{revtex4-1} %} %{article}%
\pdfoutput=1 %,floatfix,onecolumn,
\usepackage{geometry}
\usepackage{amsmath,amssymb,bm}
 \usepackage{graphicx}
 \usepackage{subfigure}
\usepackage{fancyhdr}
\usepackage{titlesec}  
\usepackage{cite}

\usepackage[
       colorlinks=true,
      filecolor=black,
       anchorcolor=blue,
      linkcolor=blue,
      citecolor=red,
      urlcolor=cyan, % url in \href{}  brown, cyan,
       linktocpage=true,
        plainpages=false,
        breaklinks=true,
            pdfstartview=FitH
           ]{hyperref}

\def\Mp{M_P}

\def\O{\Omega}
\def\L{\Lambda}

\def\d{\rm{d}}

\def\h{I}

\def\p{\partial}

\def\tg{\tilde{g}}
\def\T{\langle\mathcal{T}\rangle} %\langle\rangle

\def\w{y}
\def\nt{{\bf{n}}}
\def\at{{\bf{a}}}

\def\K{\mathcal{K}}
\def\M{\mathcal{M}}

\def\R{\mathcal{R}}

\def\H{\mathcal{H}}

\def\mS{\mathcal{S}}
\def\mM{m}
\def\dh{d}
\def\Om{\Omega_m}

\def\({\left(}
\def\){\right)}
\def\[{\left[}
\def\]{\right]}

\usepackage[symbol]{footmisc}
\newcommand{\sizef}{\footnotesize}
\title{\bf  \large Emergent Dark Universe and the Swampland Criteria} %Refined de Sitter Conjecture}

\author{\normalsize {Rong-Gen Cai$^{1,2}$}, \quad {Sunly Khimphun$^{3}$}, \quad {Bum-Hoon Lee$^{4}$}, \quad \\
\normalsize {Sichun Sun$^{5, 6}$},  ~  {Gansukh Tumurtushaa$^{7}$}, ~  {Yun-Long Zhang$^{8,9}$}\vspace{3pt}\\
\it \sizef{$^{1}$CAS Key Laboratory of Theoretical Physics, Institute of Theoretical Physics, }\\
\it \sizef{Chinese Academy of Sciences(CAS), Beijing 100190, China}\\
\it \sizef{$^{2}$School of Physical Sciences,University of Chinese Academy of Sciences, Beijing 100049, China}\\
\it \sizef{$^{3}$Graduate School of Science, Royal University of Phnom Penh, Phnom Penh, Cambodia}\\
\it \sizef{$^{4}$Center for Quantum Spacetime(CQUeST), Sogang University, Seoul 121-742, Korea}\\
\it \sizef{$^{5}$Department of Physics, Sapienza University of Rome, Rome I-00185, Italy}\\
\it \sizef{$^{6}$Department of Physics, National Taiwan University, Taipei 10617, Taiwan}\\
\it \sizef{$^{7}$IBS Center for Theoretical Physics of the Universe(CTPU),}\\
\it \sizef{Institute for Basic Science(IBS), Daejeon 34051, Korea}\\
\it \sizef{$^{8}$Center for Gravitational Physics, Yukawa Institute for Theoretical Physics(YITP)}, \\
\it \sizef{Kyoto University, Kyoto 606-8502, Japan} \\ % (CGP)
\it \sizef{$^{9}$Asia Pacific Center for Theoretical Physics (APCTP),  Pohang 790-784, Korea}  
}

\date{\sizef (\today)} 

\usepackage{fancyhdr}
\begin{document}
 
\maketitle
\thispagestyle{fancy}
%\lipsum[4-57] %\affiliation %\begin{center}\bf Abstract\end{center} The induced metric on the hypersurface has assumed to be FRW, and

\begin{abstract} 
\vspace{5pt}
We study a model of the emergent dark universe, which lives on the time-like hypersurface in a five-dimensional bulk spacetime.  The holographic fluid on the hypersurface is assumed to play the role of the dark sector, mainly including the dark energy and apparent dark matter. Based on the modified Friedmann equations, we present a Markov-Chain-Monte-Carlo analysis with the observational data, including type Ia Supernova and the direct measurement of the Hubble constant. We obtain a good fitting result and the matter component turns out to be small enough, which matches well with our theoretical assumption that only the normal matter is required. After considering the fitting parameters, an effective potential of the model with a dynamical scalar field is reconstructed. The parameters in the swampland criteria are extracted, and they satisfy the criteria at the present epoch but are in tension with the criteria if the potential is extended to the future direction. The method to reconstruct the potential is helpful to study the swampland criteria of other models without an explicit scalar field.  
 
% based on the current observation
%%JLA compilation of
%It turns out that this model satisfies the criteria at the present, but might violate them in the future. recent versions of
%Through fitting with the Type Ia Supernova, we read out the parameters in the modified Friedmann equation.   to the SNIa data with the.

~~\\ \sizef
Emails: \it ~ {cairg@itp.ac.cn};
~{kpslourk@gmail.com};
~{bhl@sogang.ac.kr}; \\ \sizef
{sichunssun@gmail.com};
{gansuhmgl@ibs.re.kr};
{yun-long.zhang@yukawa.kyoto-u.ac.jp};
\end{abstract}

%\preprint{CTPU-PTC-18-42"}

\newpage
\pagestyle{plain}

\tableofcontents
\allowdisplaybreaks
%\newpage

\section{Introduction}
%\begin{center} {\bf I. Introduction} \end{center}

The origin of dark energy and dark matter is one of the most mysterious issues in the current cosmological observations. In the Lambda-Cold-Dark-Matter (LCDM) model of the universe, roughly $95\%$ of the energy component of the current universe is invisible to us. The LCDM model is successful at the CMB and large scales, but at the galactic scale, some discrepancies were proposed \cite{Clifton:2011jh, Famaey:2011kh, Milgrom:1983ca}. Moreover, the particle dark matter is still elusive from the direct detection \cite{Undagoitia:2015gya}.  One of the alternatives to the particle dark matter is the modified gravity \cite{Clifton:2011jh, Famaey:2011kh}, which can be applied to both the inflationary era and late time acceleration of the universe. Inspired by Verlinde's recent derivation on the emergent gravity \cite{Verlinde:2016toy}, a holographic model of the Emergent Dark Universe (hEDU) has been proposed in \cite{Cai:2017asf}, where the shortened name ``hFRW'' was also used. It assumes that the stress-energy tensor of the total dark components, including dark energy and apparent dark matter, is provided by the holographic fluid on the $3+1$ dimensional Friedmann-Robertson-Walker (FRW) hypersurface, which is embedded in a $4+1$ dimensional Minkowski bulk. The Einstein field equations on the hypersurface are modified as 
\begin{align}\label{Einstein1}
 {R}_{\mu\nu} - \frac{1}{2} {R} g_{\mu\nu} 
-\frac{1}{L} ( {\K}_{\mu\nu} - {\K} g_{\mu\nu} )
  =\kappa_4\, T_{\mu\nu} \,.
\end{align}
The Einstein's constant $\kappa_4=\frac{8\pi G}{c^4}=\frac{ \hbar }{\Mp^2 c^3}$ is related to the Newton constant $G$ and the reduced Planck mass $\Mp$ in $3+1$ dimensions.  $T_{\mu\nu} $ is the stress-energy tensor of the normal matter within the standard model, $R_{\mu\nu}$ and ${\K}_{\mu\nu}$ are the intrinsic Ricci curvature and the extrinsic curvature of the hypersurface, respectively.
The length scale $L$ is assumed to be at the same order as ${c}/{H_0}$  \cite{Cai:2017asf}, where $c$ is the speed of light and $H_0$ is the Hubble constant today.  Thus, the Einstein gravity gets modified in the weak gravity region.

The modified Friedmann equation in this {hEDU} model can be derived from \eqref{Einstein1}, once taking $g_{\mu\nu}$ as the FRW metric and considering the consistent embedding into the {flat} bulk.  Based on this,  we present the Markov Chain Monte Carlo (MCMC) analysis with the Type Ia Supernova (SNIa) data and obtain a good fitting result. It turns out that, only  3\% of the component in the current universe is required to be the normal matter in this hEDU model, instead of the 30\% in the LCDM model where the dark matter has to be included.   It matches well with our theoretical assumption that only normal matter is required on the right-hand side of the modified Einstein field equations in \eqref{Einstein1}.

The emergent dark universe model will be asymptotical de Sitter (dS) in the future infinity. However, it seems difficult to construct a meta-stable dS vacuum in string theory. A group of authors \cite{Ooguri:2006in,Obied:2018sgi,Agrawal:2018own, Ooguri:2018wrx} proposed the conjecture that the scalar potential in low energy effective theory satisfies,  
\begin{align}
\text{Criterion 1}: \label{SC1}&\quad \frac{|\Delta\phi|}{\Mp} \leq d_0, \\
\text{Criterion 2}: \label{SC2}&\quad {\Mp}  {|\nabla V|} \geq c_1  {V}  \quad \text{or} \quad 
 {\Mp^2}\, {\text{min}[\nabla_i \nabla_j V]} \leq - {c_2} {V}, 
 \end{align}
over a certain range of the scalar fields and the positive constants $d_0, c_1, c_2$ are of order one $\sim\mathcal{O}(1)$, if the theory has an ultraviolet (UV) completion consistent with quantum gravity. Otherwise, the scalar potential is too flat and the theory lies in the swampland.
These conjectures constrain the possible forms of the effective scalar potentials from the top-down models, which have been studied in the inflationary era \cite{SCIF}, present dark energy dominated universe \cite{SCDE} and the effective potentials in phenomenology \cite{SCPH}. 
The swampland conjectures have been used to discuss the possible de Sitter vacua from the compactifications of string theories \cite{KKLT},
related to the Kachru-Kallosh-Linde-Trivedi (KKLT) approach \cite{Kachru:2003aw}, see also \cite{Danielsson:2018ztv} and \cite{Kallosh:2019axr}.

In our dark universe model, the dark sector arises from the holographic stress-energy tensor, which drives the expanding universe. The same stress-energy tensor can also be reconstructed from the Lagrange density of a scalar field with an effective potential. From the Friedman equation with the fitting parameters in our model, we derive the effective scalar potential numerically, such that the derivative of the potential can also be calculated.
We check the conditions in our model and comment on the swampland criteria. 
We find that the Criterion 1 in \eqref{SC1} in the hEDU model. However, near the bottom of the effective potential of hEDU model, we only have
\begin{align}\label{SC3}
{\Mp^2} {|\nabla_i \nabla_j V|} \geq  c_3  {V} ,\qquad c_3\sim \mathcal{O}(1),
\end{align}
instead of Criterion 2 in \eqref{SC2}. Especially now this condition \eqref{SC3} can include some braneworld models which are asymptotic de Sitter in the future infinity and avoid some complications at the bottom of the potential that $\nabla V=0$.

In the following section \ref{secFRW},  this dark universe model in a {flat} bulk is reviewed and the modified Friedmann equation in the present universe is derived. In section \ref{secFitting}, the parameters in the modified Friedmann equation of hEDU model are fitted with the SNIa and $H_0$ data. Based on these results, in section \ref{secSC}, the effective potential of a dynamical scalar field is reconstructed to recover the same evolution equation numerically, then the parameters in the swampland criteria can be calculated. The conclusion and discussion are summarised in section \ref{secConclusion}.

\section{Emergent Dark Universe on a Hypersurface} \label{secFRW}

We consider a $3+1$ dimensional time like hypersurface $\H$ with the induced metric $g_{\mu\nu}$ and Ricci scalar $R$, which is embedded into a $4+1$ dimensional bulk spacetime $\M$ with metric $\tg_{AB}$ and Ricci scalar $\R$. After including the Lagrangian density  ${\cal{L}}_{\mM}$ of the standard model matter on the hypersurface, we can write down the total action
\begin{align}\label{DGP}
{\mS}_{tot} & %={\mS}_{4}+{\mS}_{5} , \\ v{\mS}_{4} &
= \int_{\H} {\d}^4x \sqrt{-{g}} \,\Big( \, \frac{1}{2 \kappa_4}  R +  {\cal{L}}_{\mM} \, \Big) +{\mS}_{5}  \, ,\\
{\mS}_{5} &\equiv  \int_{\M} {\d}^{5} x \sqrt{-\tilde{g}} \,  \Big( \frac{ 1}{2 \kappa_{5}} \R \Big) 
+  \int_{\H} {\d}^{4}x \sqrt{-{g}}\,  \frac{1}{\kappa_{5}} \K  ,
\end{align}
where  $\K$ is the trace of extrinsic curvature of the hypersurface ${\H}$. The Einstein field equations on the hypersurface become \cite{Cai:2017asf},
\begin{align}
\frac{1}{\kappa_4} G_{\mu\nu}&=T_{\mu\nu}^{\mM} + \T_{\mu\nu}^{{\dh}},
\end{align}
where the Brown-York stress-energy tensor \cite{Brown:1992br} on ${\H}$ is given by
\begin{align}\label{holographicBY}
 \T_{\mu\nu}^{\dh}&\equiv -\frac{2}{\sqrt{-g}} \frac{ \delta({\mS}_5)}{\delta g^{\mu\nu}}= \frac{1}{\kappa_5}\( {{\K}_{\mu\nu}} -{\K} g_{\mu\nu}\).
\end{align}
After setting $\kappa_5= L {\kappa_4}$, we can reach the modified Einstein field equations in \eqref{Einstein1}.
Notice that in the cutoff holography on fluid/gravity duality, there is no dynamics of the induced metric on the hypersurface \cite{Bredberg:2010ky, Bredberg:2011jq,Compere:2011dx,Cai:2011xv,Cai:2014ywa}.
Although the modified Einstein field equations are related to the Dvali-Gabadadze-Porrati (DGP) braneworld models \cite{Dvali:2000hr,Deffayet:2000uy,Deffayet:2001pu}, we will give a physical interpretation from holographic scenario together with new parameters.

Considering that our universe is uniform and isotropic at large scale, we take the spatially flat {FRW} metric in $3+1$ dimensions, 
\begin{align}\label{FRW2}
{\d} s^2_4 =& g_{\mu\nu} {\d}x^{\mu} {\d}x^{\nu} = -c^2 {\d} t^2 +a(t)^2\[{\d}{r}^2 +{ r}^2 {\d}\O_{2}\].
\end{align}
The consistent embedding of this  {FRW} metric in $4+1$ dimensional {flat} spacetime has been studied in \cite{Binetruy:1999ut}, where the bulk metric in Gaussian normal coordinates is
\begin{align}\label{Gaussmetric1}
{\d} s^2_5  %=\tg_{AB} {\d}x^{A} {\d}x^{B} 
=  {\d} \w^2  - \nt(\w,t)^2 \  c^2 {\d} t^2 +\at(\w,t)^2\[{\d}{r}^2 +{ r}^2 {\d}\O_{2}\] .
\end{align}
The consistent embedding functions are solved as  \cite{Dick:2001sc, Dick:2001np, Lue:2002fe},
\begin{align}
\at(\w,t)^2 &=a(t)^2 +  \w^2  \frac{\dot{a}(t)^2}{c^2} \pm 2 \w \sqrt{a(t)^2\frac{ \dot{a}(t)^2}{c^2}+\frac{I}{L^2} } ,\label{Gat}\\
\nt(\w,t)  &=\frac{\p_t{\at}(\w,t)}{\dot{a}(t)} \, .\label{Gnt}
\end{align}
The integration constant ${I}$ is dimensionless after putting a scale factor $L^2$ in \eqref{Gat}. In the coordinates of this metric \eqref{Gaussmetric1}, the hypersurface ${\H}$ is located at $\w=0$, which is the shared boundary of the half bulk $\M_+$ for the region $\w>0$ and the half bulk $\M_-$ for the region $\w<0$.

In the spirit of the membrane paradigm \cite{Price:1986yy, Parikh:1997ma}, we remove half part of the bulk spacetime, which can be effectively replaced by the holographic stress tensor  $\T^{\dh}_{\mu\nu}$ in \eqref{holographicBY}. The dynamics of a {FRW} hypersurface which is embedded into the higher dimensional {flat} spacetime has been studied in \cite{Cai:2017asf}. With the bulk metric \eqref{Gaussmetric1}, the energy density and pressure in  $\T^{\dh}_{\mu\nu}$ are calculated to be
\begin{align}
\rho_{\dh}(t)  & =\rho_c \sqrt{\Omega_\L}\sqrt{ \frac{H(t)^2}{H_0^2} + \frac{\Omega_\h}{a(t)^4} } , \label{rhoH} \\
p_{\dh}(t) &=- \frac{\dot\rho_{\dh} }{3 H(t)} -\rho_{\dh} \, ,\label{pH}
\end{align}
where the critical density and other parameters are given by 
\begin{align}\label{Friedmann2}
\rho_{c} & =  \frac{3 H_0^2 M_P^2 }{\hbar c},\quad \O_{\L} =\frac{c^2}{L^2 H_0^2}, \quad  \O_{I}\equiv \frac{\h  c^2}{L^2 H_0^2}.
\end{align}
The effective cosmological constant in the future infinity turns out to be $\Lambda= \frac{3}{L^2}$  \cite{Cai:2017asf}.
The modified Friedmann equation becomes 
\begin{align}\label{Friedmann2}
H(t)^2 &= \frac{\kappa_4 c^4 }{3} \[ \rho_{\mM} (t) + \rho_{\dh}(t)\].
\end{align}

Plugging \eqref{rhoH} into \eqref{Friedmann2} and considering the relation between the redshift $z$ and the scale factor via $a(t)/a(t_0)=1/(1+z)$, we arrive at the normalized Hubble parameters $ {H(z)}/{H_0}$ in terms of the redshift $z$, which is the modified Friedmann equation in the {hEDU} 
\begin{align} 
\frac{H(z)^2 }{H_0^2 } & = {\frac{\O_{\L}}{2}+ {{\Om}}{(1+z)^3} 
+  \frac{\O_{\L} }{2} \sqrt{ 1 +  \frac{4}{\O_{\L}}\Big[   {\Om} {(1+z)^3} +   \O_I (1+z)^4  \Big]}} .
 \label{hFRWfriedmann}
\end{align}
Taking \eqref{hFRWfriedmann} at $z=0$ with $H_0\equiv H(z)|_{z=0}$, we have the relation between different components
\begin{align} \label{OmegaL}
1&={\Om}+\sqrt{\O_\L(1 +\O_I)} ~\Rightarrow ~\O_\L=\frac{(1-{\Om})^2}{(1 +\O_I) }.
\end{align}
Notice here that by setting $\O_{I}=0$, we can recover the usual Friedmann equation of the self-accelerating branch of the DGP braneworld model (sDGP). When $\O_{I}\ll 1$, the behavior of this term is more like the dark radiation \cite{Mukohyama:1999qx}. In this holographic model of the emergent dark universe ({hEDU}), $\O_I$ turns out not to be so small, such that the whole dark sector, including the dark energy and apparent dark matter, is expected to be included in the holographic dark fluid \cite{Cai:2017asf}.

\section{Fitting Parameters with the SNIa and $H_0$ data} \label{secFitting}

In this section, we give constraints on model parameters in the modified Friedmann equation~(\ref{hFRWfriedmann}) in light of observational data. We will also compare the result with the flat LCDM model, where the standard Friedmann equation is
\begin{align}
\text{LCDM}:
\quad  \frac{H(z)^2 }{H_0^2 } & = { {\O_{\L}} + {\Om}{(1+z)^3}  },\qquad  \O_\L= 1- {\Om}.
 \label{LCDM friedmann}
\end{align}
We employ the Markov-chain Monte Carlo (MCMC) together with the observational data and the statistical methods. In particular, in our analysis, we use Type Ia supernovae (SNIa) observational data and the direct measurement of Hubble constant $H_0$. There are different types of SNIa datasets compiled by different groups. Thus, in this work, we consider Joint Light-curve Analysis (JLA) dataset~\cite{Betoule:2014frx}, which consists of 740 supernovae. JLA data includes 118 supernovae at $0<z<0.1$ from several low-redshift samples, 374 supernovae at $0.03<z<0.4$ from the Sloan Digital Sky Survey (SDSS) SN search, 239 supernovae at $0.1<z<1.1$ from the Supernova Legacy Survey (SNLS) observation, and 9 supernovae at $0.8<z<1.3$ from the Hubble Space Telescope (HST) measurements. In addition, we also use $H_0$ data derived from a reanalysis of Cepheid data~\cite{Riess:2011abc, Riess:2016jrr} to constrain our model.

\begin{figure}[h] %[htbp]%
\centering  %\subfigure[hFRW]
{\includegraphics[scale=0.32]{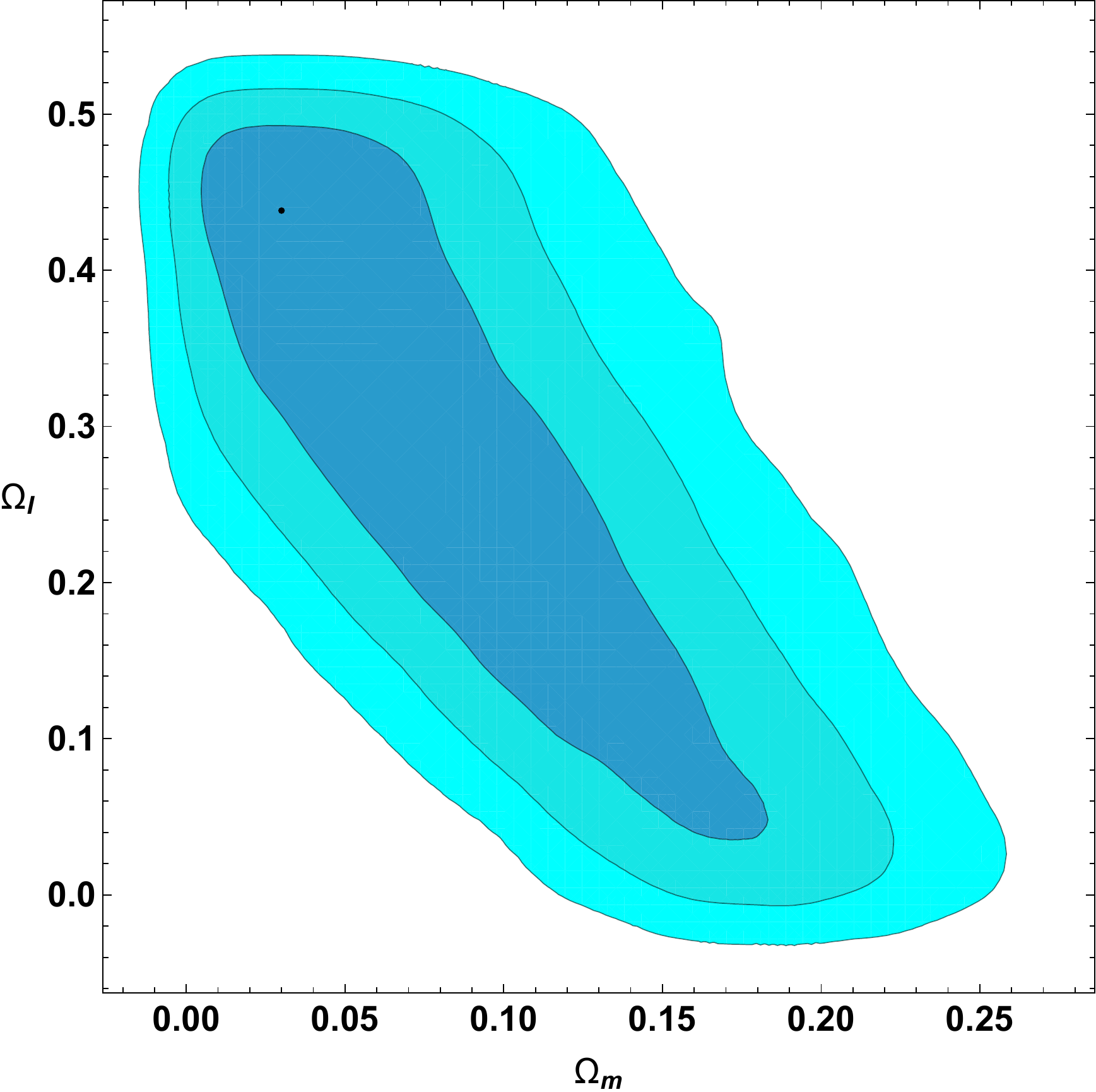} } %0.23
\qquad {\includegraphics[scale=0.32]{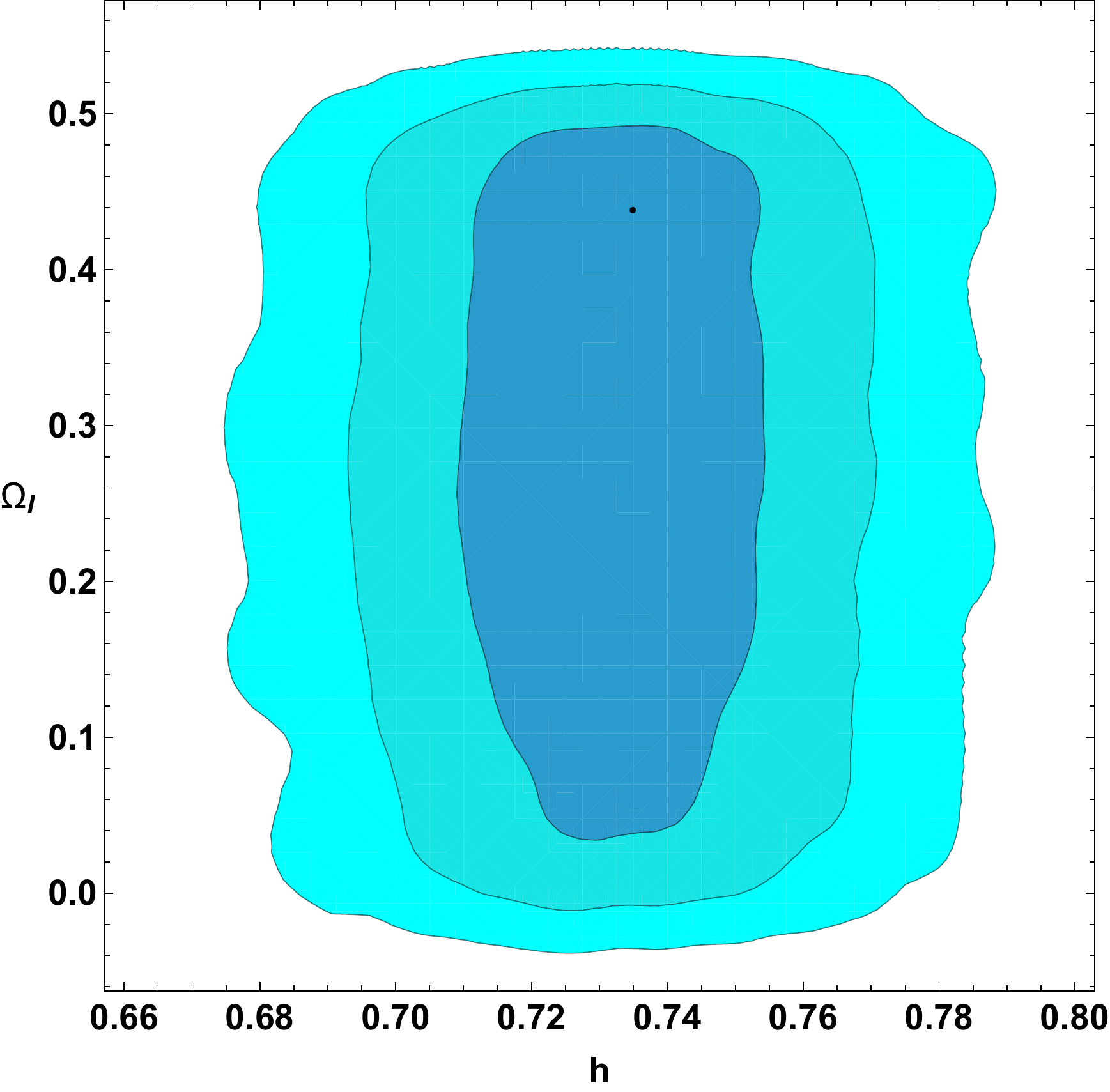}} \vspace{15pt} \\  %0.23
{\includegraphics[scale=0.32]{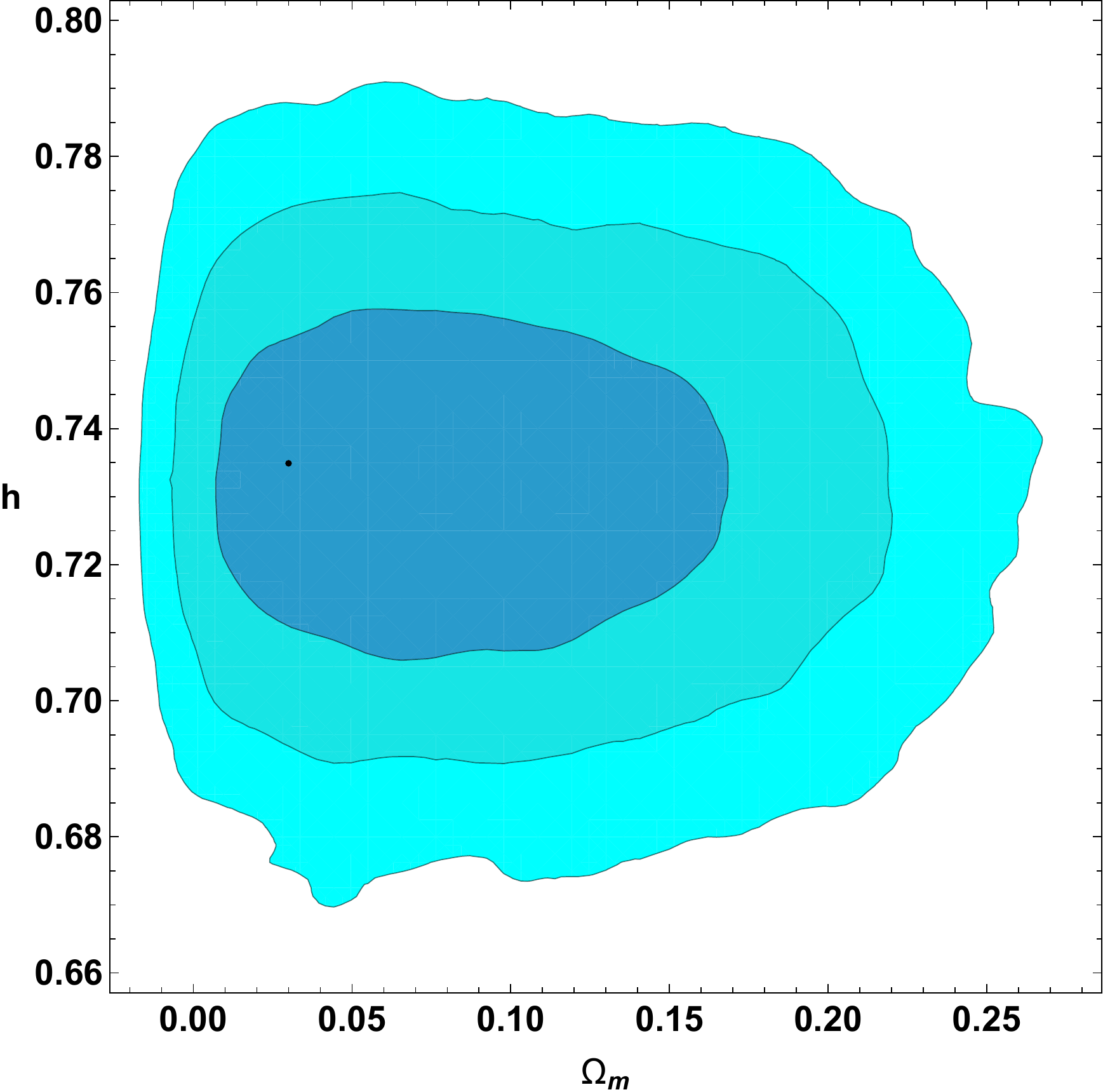} } % 0.235
\qquad ~  {\includegraphics[scale=0.32]{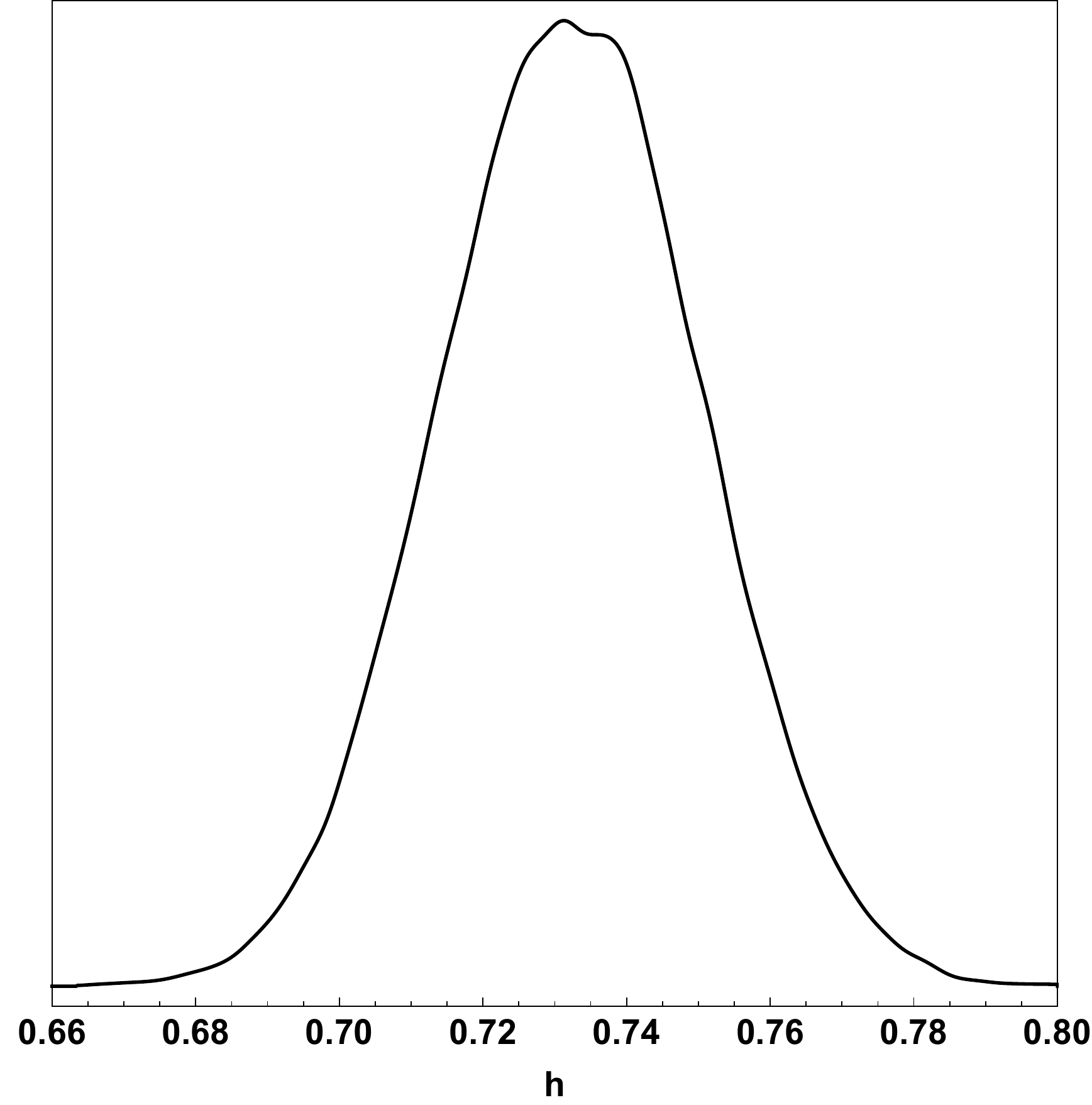} } %0.223
\caption{The $68.3\%$, $95.4\%$, and $99.7\%$ confidence contours for various parameter combinations. $\Omega_m$, $\Omega_I$, $h=H_0/(100\,\text{km}\,\text{s}^{-1}\,\text{Mpc}^{-1})$ and 1D marginalized likelihood for $h$. The best fit values are at $\Omega_m=0.0299$, $\Omega_I = 0.4382$ and $h= 0.7349$.}
\end{figure}

We use the $\chi^2$ statistics to fit the cosmological models to observational data. The $\chi^2$ function is given by 
\begin{equation}\label{eq:chi2}
\chi_\theta^2=\frac{(\theta_\text{th}-\theta_\text{obs})^2}{\sigma_\theta^2}\,,
\end{equation}
where $\theta_\text{obs}$ is the measured value by observations, whereas $\theta_\text{th}$ is the predicted value by theory, and $\sigma_\theta$ is the standard deviation. The total $\chi^2$ is the sum of all $\chi^2_\theta$,
\begin{equation}\label{eq:totalchi2}
\chi^2=\sum_\theta \chi^2_\theta\,.
\end{equation}

As we mentioned above, the observational data we use in this work include the JLA sample of SNIa observation and the direct measurements of the Hubble constant $H_0$ from HST. The total $\chi^2$ is written as $\chi^2=\chi^2_\text{SNIa}+\chi^2_{H_0}$. Therefore, in the following, we will include both the SNIa and $H_0$ data into the $\chi^2$ statistics.

We use the JLA compilation of SNIa. The JLA compilation includes 740 Ia supernovae in range of $z \in [0.01, 1.3]$.  According to the observational point of view, we need the distance modulus $\mu$, which is often used to construct the $\chi^2$ function of SNIa, and the redshift $z$ to the SNIa to fit the data. the theoretical value of distance modulus can be computed as 
\begin{eqnarray}
\mu_\text{th}=5\log_{10}\left[\frac{d_L(z)}{\text{Mpc}}\right]+25\,,
\end{eqnarray}
where $d_L$ is the luminosity distance predicted from theory. The luminosity distance $d_L$ is given by
\begin{eqnarray}
d_L(z)= (1+z) r(z)\,,
\end{eqnarray}
where
\begin{eqnarray}
r(z)= H_0^{-1}|\Omega_K|^{-\frac12}\sinh\left[|\Omega_K|^{\frac12}\int^z_0\frac{dz'}{E(z')} \right]\,.
\end{eqnarray}
Here, $H_0=100 h\, \text{km}\, \text{s}^{-1}\, \text{Mpc}^{-1}$ is the Hubble constant, $E(z)$ is the reduced Hubble parameter and is defined as $E(z)\equiv H(z)/H_0$, and $\sinh(x)=\sin(x), x, \sinh(x)$ for $\Omega_K<0$, $\Omega_K=0$, and $\Omega_K>0$, respectively. 

The observed value of the distance modulus is given as 
\begin{eqnarray}
\mu_\text{obs} = m_B^*-M_B+\alpha  \, X_1 +\beta \, \mathcal{C}\,,
\end{eqnarray}
where $m_B^*$ is the observed peak magnitude in the rest-frame of $B$ band, $X_1$ describes the time stretching of light-curve, and $\mathcal{C}$ describes the SN color at maximum brightness. As we mentioned above, the JLA data includes 740 SNIa; for each SNIa, the observed values of $m_B^*$, $X_1$, and $\mathcal{C}$ are given in reference~ \cite{Betoule:2014frx}. The $\chi^2$ function for JLA observation can be written as 
\begin{equation}\label{eq:chi2SNIa}
\chi^2_\text{SNIa} = (\mu_\text{obs}-\mu_\text{th})^\dagger \text{Cov}_\text{SNIa}^{-1} (\mu_\text{obs}-\mu_\text{th})\,,
\end{equation}
where $\text{Cov}_\text{SNIa}$ is the covariance matrix of the JLA observation.

For the $H_0$ measurement, we use the result of direct measurement of Hubble constant, given by Riess {\emph et al.}~\cite{Riess:2016jrr}, $H_0=73.24\pm1.74 \text{km}\, \text{s}^{-1} \text{Mpc}^{-1}$, which is derived from a re-analysis of Cepheid data. However, this measurement is in tension with Planck data~\cite{Adam:2015rua}. The $\chi^2$ function for the $H_0$ measurement is
\begin{equation}\label{eq:chi2H0}
\chi^2_{H_0} = \left( \frac{h-0.7324}{0.0174}\right)^2\,.
\end{equation}

If we compare our model with the LCDM model, $\chi^2$ cannot  make fair comparison, for them having different numbers of free parameters, because a model with more parameters has more tendency to have a lower value of $\chi^2$. Thus, to make a fair comparison, we apply the Akaike information criterion (AIC)~\cite{AIC} and Bayesian information criterion (BIC)~\cite{BIC} to do analysis. The AIC and BIC are defined as $ \text{AIC} \equiv -2 \ln \mathcal{L}_\text{max} + 2k$ and $\text{BIC} \equiv -2\ln \mathcal{L}_\text{max} + k \ln N$, respectively, where $\mathcal{L}_\text{max}$ is the maximum likelihood, $k$ is the number of parameters, and $N$ is number of data points used in the model-data fit. For Gaussian errors, one can use $\chi^2_\text{min}=-2\ln \mathcal{L}_\text{max}$.

\begin{table}[h!]
\centering  %\label{tab1} %$\Omega_b h^2$   & $0.0986\pm0.2152$      & ---   \\ 
\begin{tabular}{c c c c c c c c c c}
 \hline\hline 
 Parameters           &\vline&  LCDM               &\vline&  {hEDU}    \\ [0.5ex] \hline
 $h$                           &\vline& $0.7330\pm0.0180$     &\vline& $0.7349\pm 0.0179$ \\  
 $\Omega_m$           &\vline& $0.2969\pm 0.0352$    &\vline& $0.0299\pm 0.0515$   \\  
$\Omega_I$             &\vline& ---                                 &\vline& $0.4382\pm 0.1317$  \\  \hline
$\alpha$                  &\vline&  $0.1403 \pm 0.0068$       &\vline& $0.1409\pm 0.0068$ \\ 
$\beta$                    &\vline& $3.1081\pm 0.0892$      &\vline& $3.1144\pm 0.0896$  \\   \hline
$\chi^2_\text{min}$  &\vline& 695.063                       &\vline& 694.321   \\  
$\Delta$AIC             &\vline& 0                               &\vline& 1.258  \\  
$\Delta$BIC             &\vline& 0                                &\vline& 5.866  \\  
 \hline\hline
\end{tabular}
\caption{Fitting values and uncertainties of the cosmological parameters. \label{tab1}} % Results of 
\end{table}
 
We introduce AIC and BIC statistics for the sake of comparing different models due to the different free parameters. Obviously, a model with a smaller AIC value means a better model in terms of data fitting, while a smaller BIC value implies that such a model is economically favorable if further data points are implemented. In our analysis, we use LCDM as a reference model, for such model is currently the best data-fitting model among all existing ones; hence, for our analysis, we need to pay more attention to the relative values of AIC and BIC as $\Delta \text{AIC}=\text{AIC}_\text{hEDU}-\Delta \text{AIC}_{{\text{LCDM}}}$ and  $\Delta \text{BIC}=\text{BIC}_\text{HPU}-\Delta \text{BIC}_{{\text{LCDM}}}$, respectively. Therefore, we need to calculate $\Delta \text{AIC}=\Delta \chi^2_{\text{min}}-2\Delta k$ and $\Delta \text{BIC}=\Delta \chi^2_\text{min}-\Delta k \text{ln}N$. It is worth noticing that, in terms of data fitting, the model with $0<\Delta \text{AIC}<2$ have a substantial support; the models with $4<\Delta \text{AIC}<7$ have considerably less support, and the models with $\Delta \text{AIC}>10$  have essentially no support, with respect to the reference model. Concerning the BIC, the relative value $\Delta \text{BIC}=\text{BIC}_\text{hEDU}-\text{BIC}_{\Lambda\text{CDM}}$ provides the following situations. The model with $\Delta\text{BIC}\leq 2$ indicates that the comparison model is consistent with the reference model. The models with $2\leq \Delta\text{BIC}\leq 6$ indicates the positive evidence against the comparison model, whereas for $\Delta \text{BIC}\geq 10$ such evidence  becomes strong. As the result shown, according to  $\Delta \text{AIC}=1.258$, our model fits well with the observational data.
However, $\Delta\text{BIC}=5.866$ indicates that if more data will be used, $\Delta$AIC between the two models might be, in some extent, increasing, so only the future data can tell us more about how well theses models relatively fit the observational data.

An overall presentation of constraints is listed in Table~\ref{tab1} for our model. The table contains the fitting parameters, including the intrinsic values ($\alpha, \beta$) of JLA, and goodness of fit statistics ($\chi_\text{min}^2$) for our model. For comparison we additionally provide the results of the usual LCDM model of cosmology. From the Table~\ref{tab1}, one can see that $H_0 = 73.49\pm 1.7998\,\text{km}\, \text{s}^{-1} \text{Mpc}^{-1}$ is the value closer to that obtained from the local measurement~\cite{Riess:2016jrr}. 
Significantly, the matter component $\O_m$ in {hEDU} model turns out to be very small, compared with the  $\O_m$ in LCDM. It matches well with our theoretical assumption in section \ref{secFRW}, that only the normal matter is required in the hEDU model.  In the next section, based on these parameters from Table~\ref{tab1}, we will recover an effective potential with the dynamical scalar field.

\section{Checking on the Swampland Criteria}\label{secSC} %Swampland Criterion and  de Sitter Conjucture
%\vspace{10pt}\begin{center} {\bf IV. The Parameters in Swampland Criterion}\label{SecV} \end{center}
%\int  {\d}^4x \sqrt{-{g}} \,R +
% It has been suggested that if an effective field theory can be embedded consistently in quantum gravity, it has to satisfy two criteria \cite{Ooguri:2006in,Obied:2018sgi,Agrawal:2018own,Ooguri:2018wrx}

%The effective field theory admits solutions modeling an accelerated universe relevant for dark energy. However, these solutions have to satisfy certain criteria in order not to end up in the Swampland.  
We can write the effective field theory of one dynamical scalar field for the late-time accelerating universe in the following action,
\begin{align}\label{scalarV}
{\mS}_{tot}&= \int  {\d}^4x \sqrt{-{g}} \Big[\frac{1}{2 \kappa_4} R + {\cal{L}}_{\mM} -\frac{1}{2} (\partial \phi)^2 - V(\phi)\Big]  .
\end{align}
 The swampland criteria \eqref{SC1} and \eqref{SC2} on an effective field theory which is supposed to be consistent with a theory of quantum gravity were reviewed in the introduction. 
 %First, the range traversed by a scalar field is bounded by $|\Delta \phi |/\Mp < d_0 \sim \mathcal{O} (1)$ \cite{Ooguri:2006in}, where $\Mp$ is the reduced Planck units .
%Second, the derivative of the scalar-field potential has to satisfy the lower bound $\Mp |\nabla V |  \geq c_1 V \sim \mathcal{O} (1)$ \cite{Obied:2018sgi,Agrawal:2018own}, or the minimum of the second derivative of the scalar-field potential has to satisfy the bound $\Mp^2(\text{min}[ \nabla_i\nabla_j V] )  \leq- c_2 V\sim \mathcal{O} (1)$ \cite{Ooguri:2018wrx}.
%{\cm If the field traverses a larger distance, then one leaves the domain of validity of the effective field theory (new string states become massless). Within a domain over which the field evolves, the second condition then needs to be satisfied. The second of these two criteria will be primarily relevant and interesting in view of dark-energy applications.  }
%Since we are interested in the dark energy dominated epoch we will assume $w_\text{matter} \sim 0$\cite{Agrawal:2018own}. 

\begin{figure}[htbp!]
\centering 
{\includegraphics[scale=0.44]{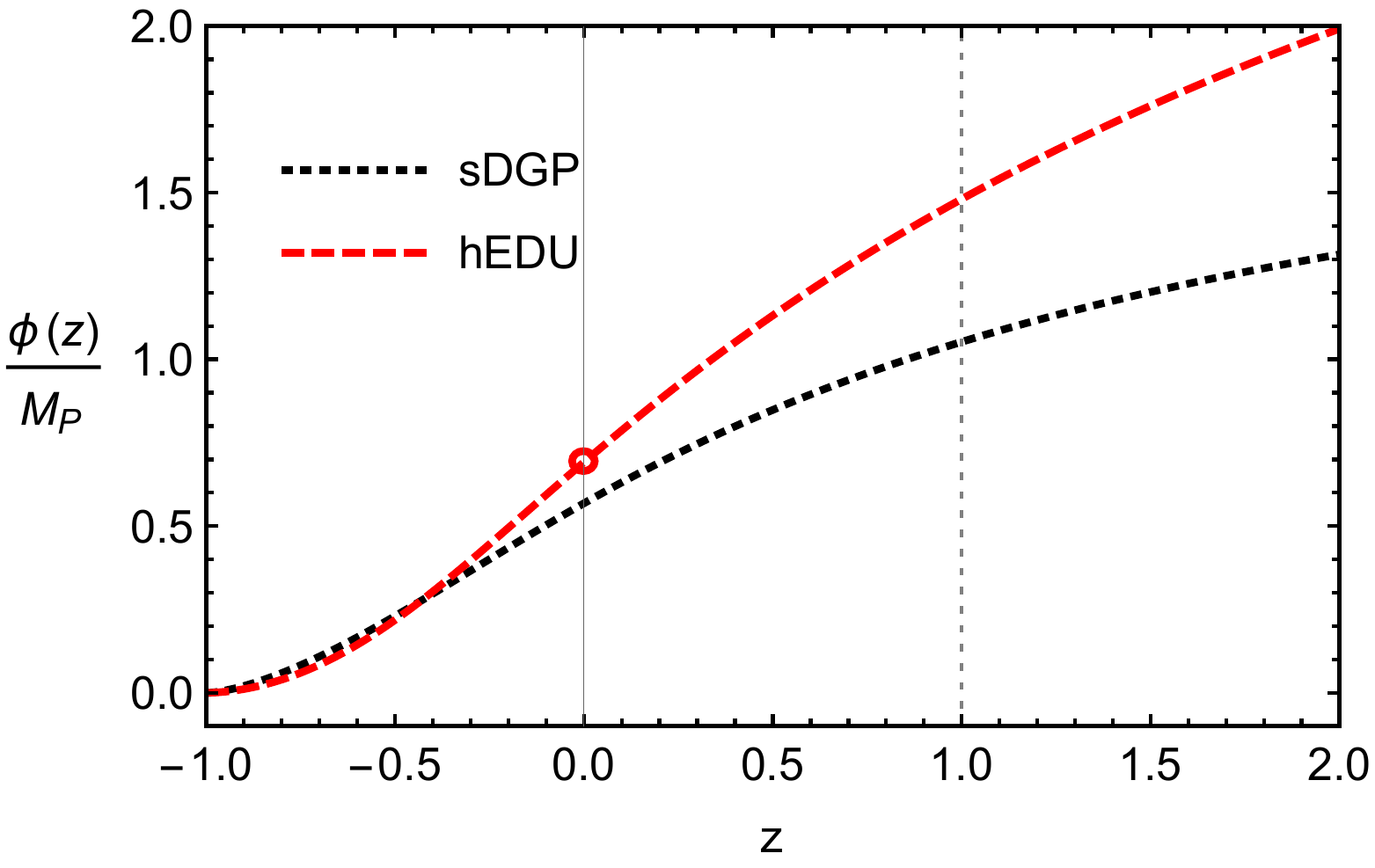}}\quad
%\end{figure}\begin{figure}[htbp!]
%\centering 
{\includegraphics[scale=0.45]{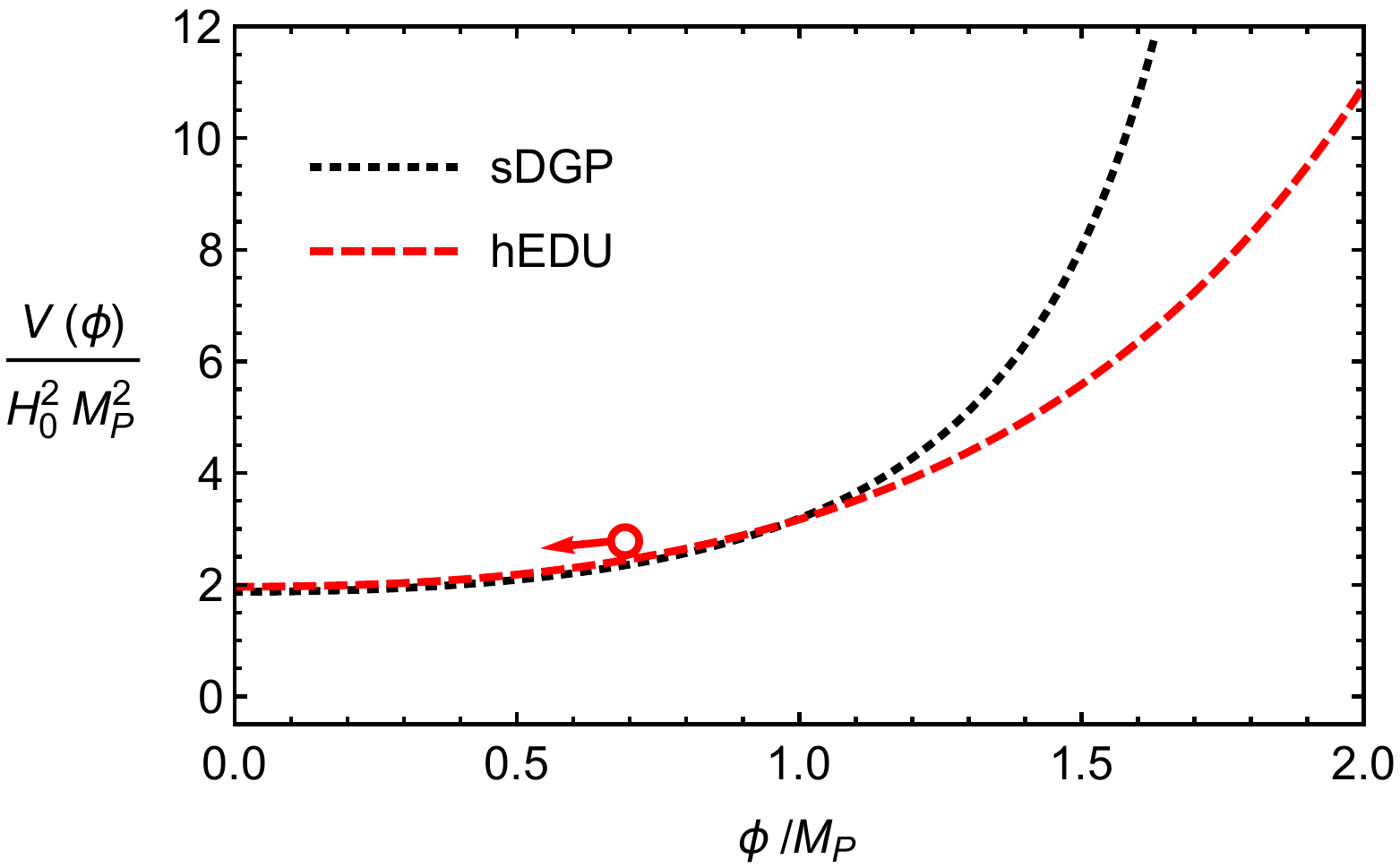} }
\caption{Left: The effective scalar field $\phi(z)$ in terms of the redshift $z$, which is related to the swampland criterion 1 in \eqref{SC1}; Right: The shape of the effective potential $V(\phi)$ in terms of $\phi$. The red circle indicates the present value of $\phi(z)|_{z=0}\simeq 0.65 \Mp$ for the {hEDU} model, and the arrow indicates the direction for the future. The parameters in the Friedmann equation \eqref{hFRWfriedmann} are taken from Table. \ref{tab2}. % \eqref{psDGP} for sDGP model and \eqref{phFRW} for {hEDU} model.
\label{figphi}} 
\end{figure}

In the hEDU model, the holographic dark fluid in \eqref{holographicBY} is assumed to be the pure gravitational effects, which can be considered as the dynamics of an effective vacuum.
So is there an effective potential of the dynamical scalar field in \eqref{scalarV}, which can recover the same effects?  From the holographic energy density \eqref{rhoH} and pressure \eqref{pH}, comparing with the energy density $\rho_{\phi}=\frac{\dot{\phi}^2}{2}+V(\phi)$ and pressure $p_{\phi}=\frac{\dot{\phi}^2}{2}-V(\phi)$ of the scalar field in \eqref{scalarV}, we can reconstruct the effective potential and the scalar field of the holographic dark fluid, which satisfies
\begin{align}
V[\phi(t)]&= \frac{1}{2} \[ \rho_{\dh}(t) - p_{\dh}(t) \] , \label{rhoH0} \\
{\dot\phi(t)} &= - \sqrt{\rho_{\dh}(t)+ p_{\dh}(t)} \, .\label{pH0}
\end{align}
For convenience, we have chosen the negative sign in \eqref{pH0}.
Taking the parameters  from Table~\ref{tab1}, we can numerically plot the $\phi(z)$ and $V(\phi)$ in Figure \ref{figphi},
by using the relation $\frac{\d t }{\d z}=\frac{-1}{(1+z)H(z)}$. 
From the modified Friedmann equation \eqref{hFRWfriedmann}, we have seen that sDGP is a special case of hEDU when $\Omega_I=0$.
Thus, in this section, we will choose the sDGP as a reference model of hEDU, along with the following parameters.
%\begin{align}  \text{sDGP}: \quad  &{\Om}\simeq 0.21, \quad \O_I\simeq 0,\qquad  \O_\L \simeq 0.62, \label{psDGP}\\
%\text{hEDU}:\quad  & {\Om}\simeq 0.03, \quad  \O_I\simeq 0.44,  \quad  \O_\L \simeq 0.65.\label{phFRW}
%\end{align}
\begin{table}[h!]
\centering  %\label{tab1}
\begin{tabular}{c c c c c c c c c c}
 \hline\hline 
 Models      &\vline&  $\Omega_m$  &\vline& $\Omega_I$ &\vline& $\Omega_\Lambda$  \\ [0.5ex] \hline
 {sDGP}     &\vline& $0.21$              &\vline& $0$          &\vline&   $0.62$   \\  
{hEDU}      &\vline & $0.03$                      &\vline& $0.44$         &\vline&   $0.65$  \\   
 \hline\hline
\end{tabular}
\caption{The input parameters of the models in section \ref{secSC}. The relation $1 ={\Om}+\sqrt{\O_\L(1 +\O_I)}$  in \eqref{OmegaL} is used to obtain $\Omega_\Lambda$.\label{tab2}}
\end{table}

The values in the sDGP model are taken from the reference \cite{Lue:2005ya}, and those values in the {hEDU} model are taken from Table \ref{tab1}.

In the figure of $\phi(z)/\Mp$ in terms of the redshift parameter $z$, the zero of $\phi(z)$ is chosen to be at the future infinity that $\phi(z)|_{z\to -1}=0$. It is clear to see that $|\Delta \phi| \sim |\phi(1)|$ is of order 1 in both models, at the dark energy dominated region from $z\simeq1$ to $z\simeq-1$.
Thus, the first swampland conjecture in \eqref{SC1} is satisfied in the present universe for both models.

Notice that in the region $0 \lesssim \frac{\phi}{\Mp}  \lesssim 1$,  the effective potentials in Figure \ref{figphi} can be fitted well with the polynomial formula,
\begin{align}\label{potential1}
 \frac{{V(\phi)}}{H_0^2 {\Mp^2} } =\frac{ \Lambda_0}{H_0^2}\,  + \frac{h_2 }{2}  \frac{\phi^2}{\Mp^2} + \frac{h_3}{3!}  \frac{\phi^3}{\Mp^3}  + \frac{h_4}{4!}   \frac{\phi^4}{\Mp^4}   +\cdots .
 \end{align}
Where $\Lambda_0= {3\Omega_\Lambda H_0^2}/{c^2}$ is the effective cosmological constant at the future infinity \cite{Cai:2017asf} and it can be calculated from $\Omega_\Lambda$ in Table \ref{tab2}.
The linear term $h_1 \frac{\phi}{\Mp}$ is dropped because we have $V'(\phi)|_{\phi\to 0}=0$. 
The fitting parameters $h_2, h_3, h_4$ are listed in Table \ref{tab3}, where $h_2>0$ implies that the effective mass of the scalar field $\phi$ is positive. 
\begin{table}[h!]
\centering  %\label{tab1}
\begin{tabular}{c c c c c c c c c c}
 \hline\hline 
 Models      &\vline&  $h_2$  &\vline& $h_3$ &\vline&  $h_4$  \\ [0.5ex] \hline
 {sDGP}     &\vline& $1.73$   &\vline& $-2.42$ &\vline& $21.0$   \\  
{hEDU}      &\vline & $1.32$  &\vline& $~2.20$  &\vline& $4.20$  \\   
 \hline\hline
\end{tabular}
\caption{The fitting parameters in the polynomial formula of the potential in \eqref{potential1}.\label{tab3}}
\end{table}

\begin{figure}[htbp!]
\centering 
\subfigure{\includegraphics[scale=0.45]{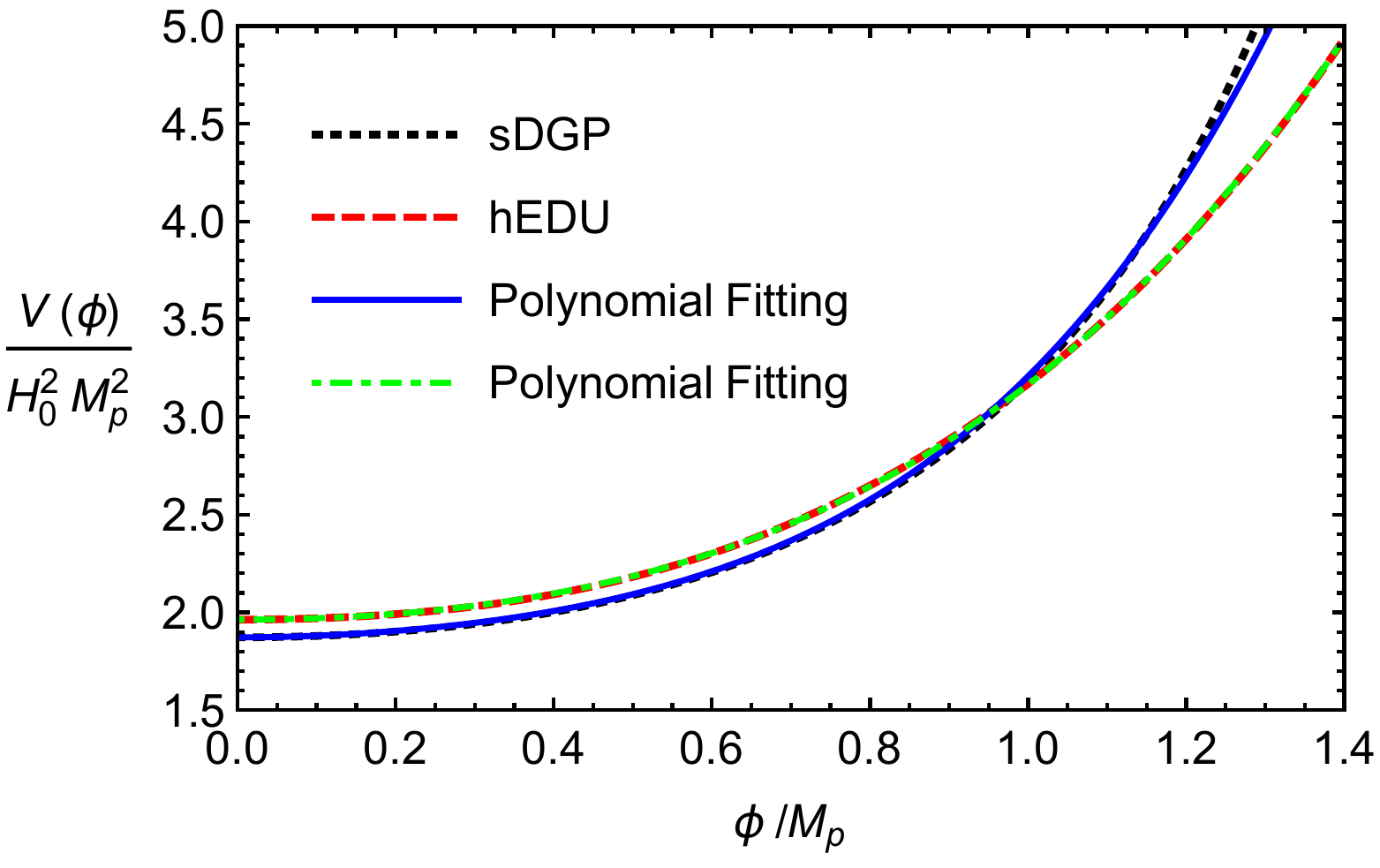} }\quad 
\subfigure{\includegraphics[scale=0.44]{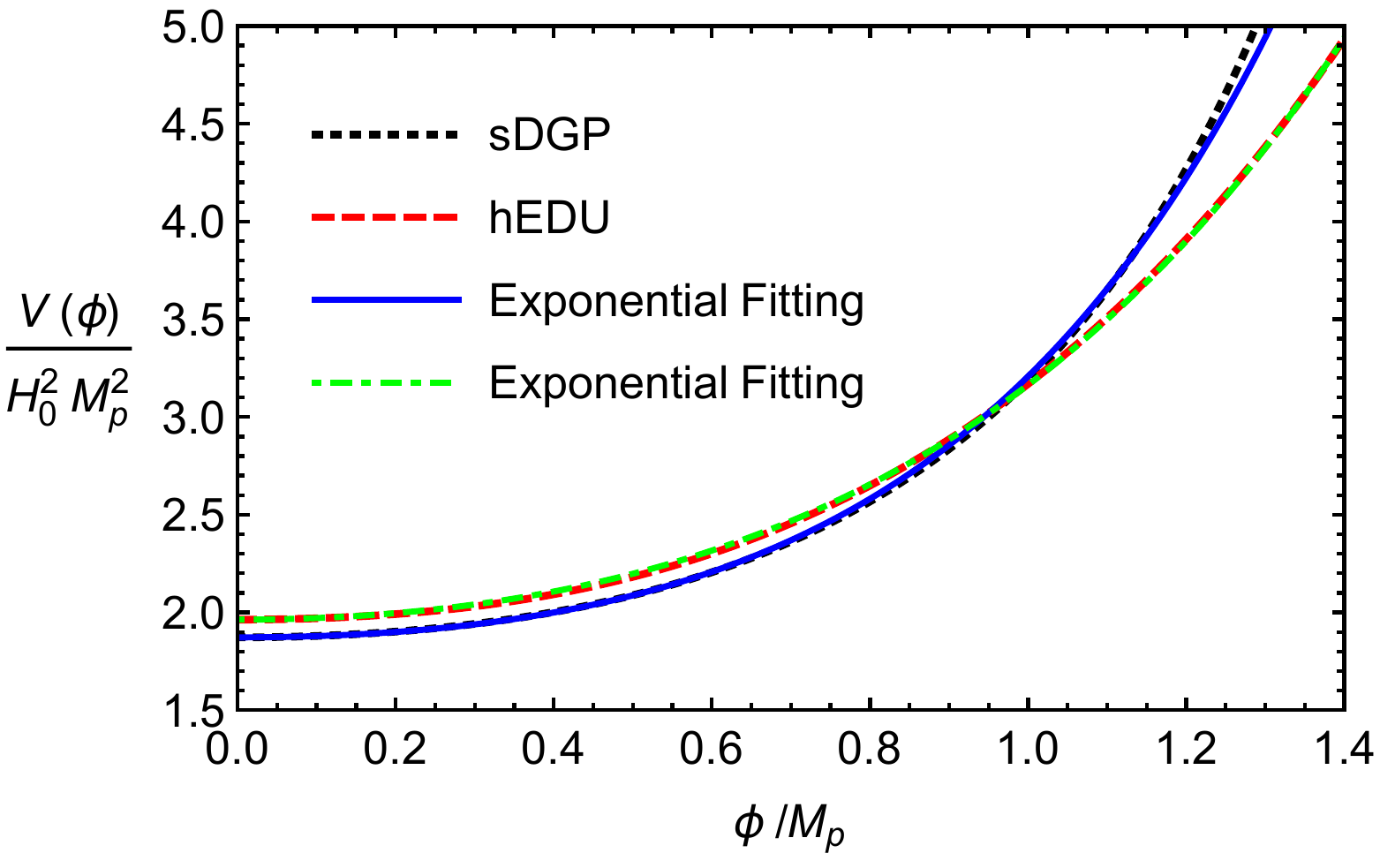} }
\caption{Left: The polynomial fittings with the potential in \eqref{potential1}, the fitted parameters are listed in Table \ref{tab3}.
Right: The exponential fittings with the potential in \eqref{potential2}, the fitted parameters are listed in Table \ref{tab4}. } 
\end{figure}

Intriguingly, the potentials can also be fitted quite well with two parameters $\lambda_{^+}$ and $\lambda_{^-}$ in the exponential formula,
\begin{align}\label{potential2}
\frac{V(\phi)}{\Mp^2}=\frac{\Lambda_0}{\lambda_{^+} +\lambda_{^-}}\Big(  {\lambda_{^+}} e^{- \lambda_{^-} \frac{\phi}{\Mp}}+ {\lambda_{^-}} e^{\lambda_{^+} \frac{ \phi}{\Mp}} \Big) .
 \end{align} 
The ansatz satisfies $V(\phi)|_{\phi\to 0}=\Lambda_0 \Mp^2$ and $V'(\phi)|_{\phi\to 0}=0$ automatically, and the fitting parameters are listed in Table \ref{tab4}. It is reasonable as there are only two free parameters $\Omega_m$ and $\Omega_I$ in the hEDU model \eqref{hFRWfriedmann}, with the relation in \eqref{OmegaL}. It is also interesting to relate the effective potential to some top down models in \cite{Obied:2018sgi}.
\begin{table}[h!]
\centering  %\label{tab1}
\begin{tabular}{c c c c c c c c c c}
 \hline\hline 
 Models      &\vline&  $\Lambda_0/H_0^2$  &\vline& $\lambda_{^+}$ &\vline& $\lambda_{^-}$  \\ [0.5ex] \hline
 {sDGP}     &\vline& $1.87$   &\vline& $2.19$ &\vline& $0.29$   \\  
{hEDU}      &\vline & $1.96$  &\vline& $1.54$  &\vline& $0.51$  \\   
 \hline\hline
\end{tabular}
\caption{The fitting parameters in the exponential formula of the potential in \eqref{potential2}.\label{tab4}}
\end{table}

\begin{figure}[htbp!]
\centering 
\subfigure{\includegraphics[scale=0.45]{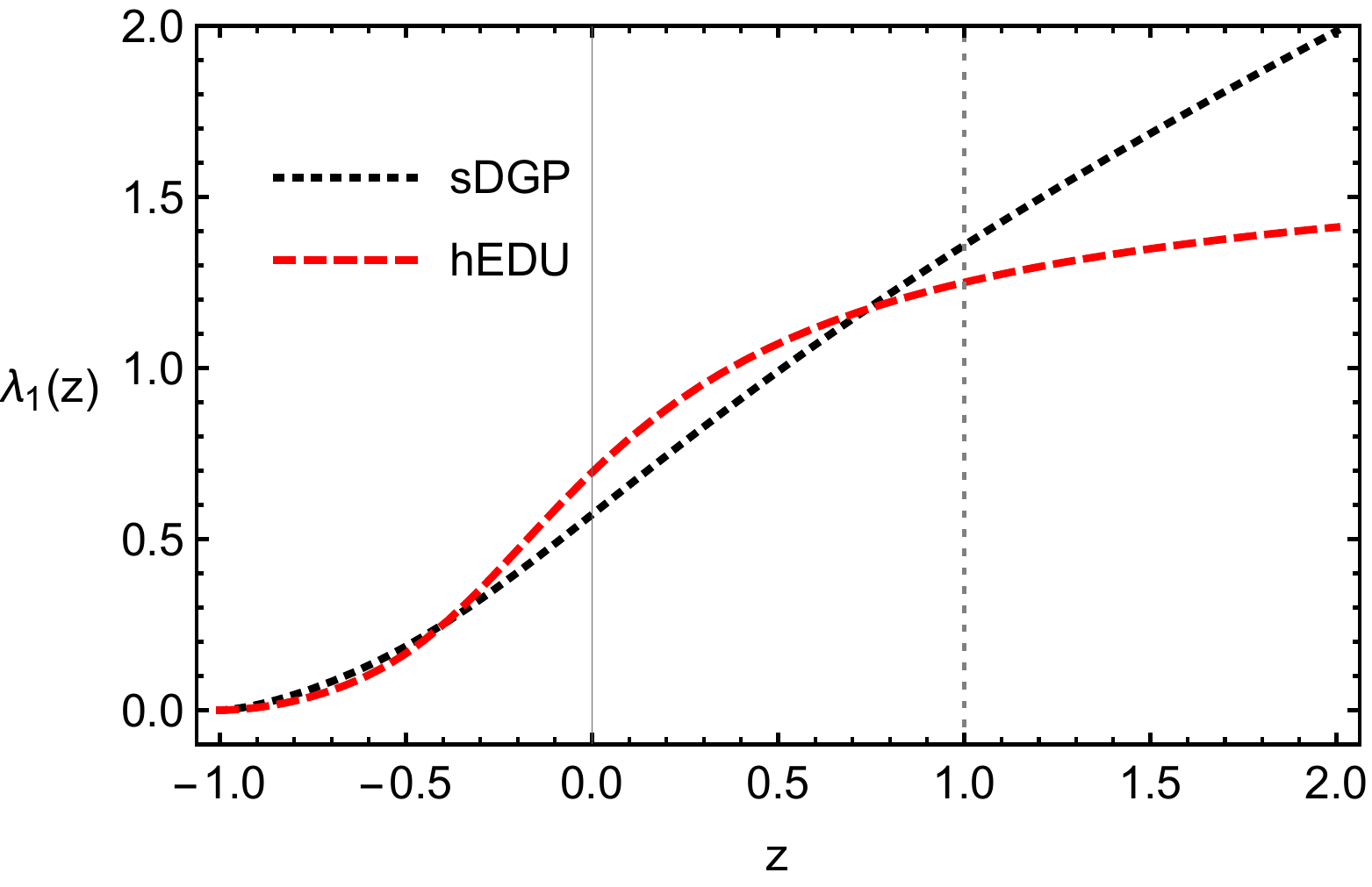} }\quad 
\subfigure{\includegraphics[scale=0.44]{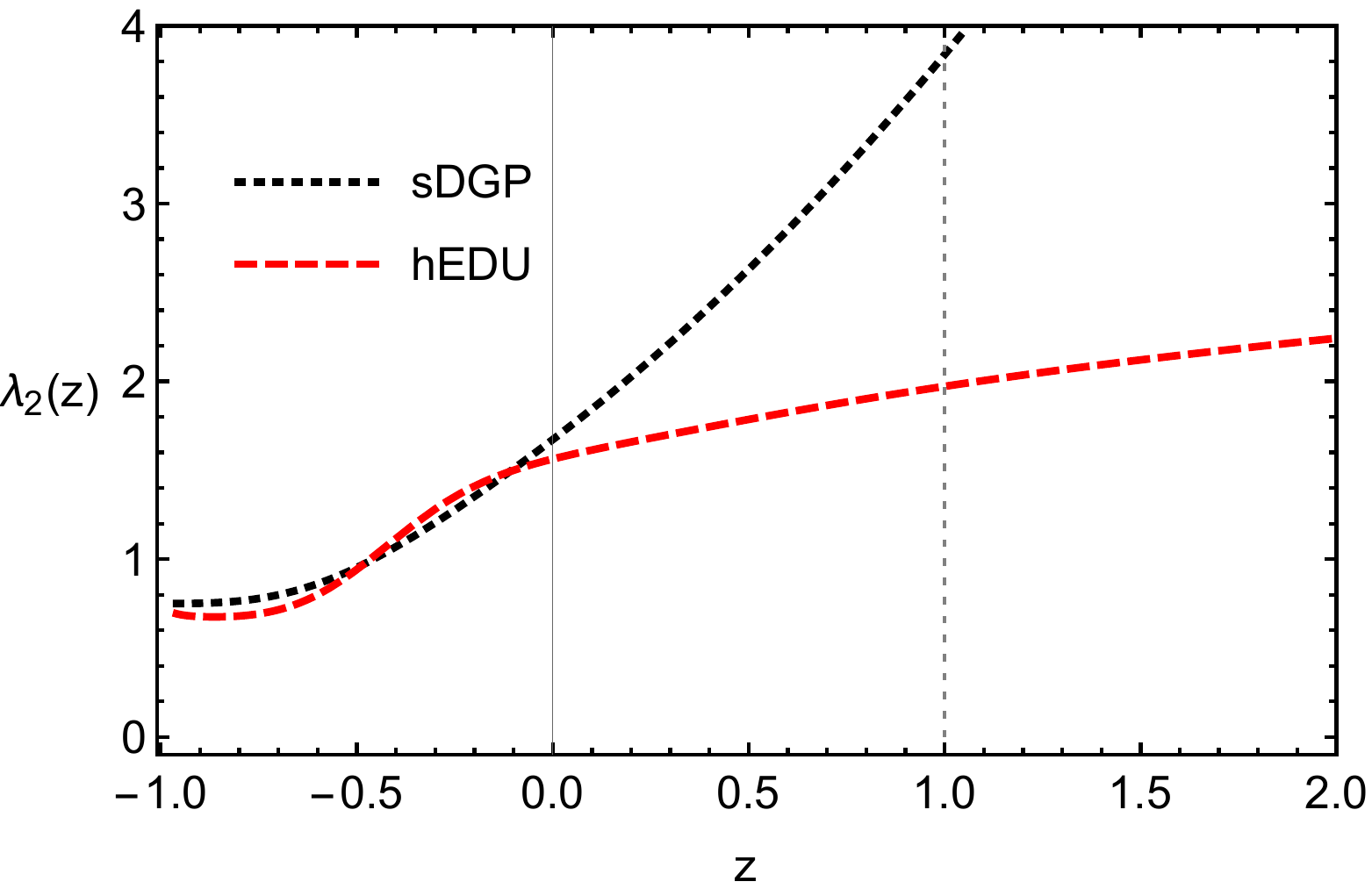} }
\caption{The parameters $\lambda_1\equiv \Mp \frac{{V'} }{V}$ in \eqref{lambda1} and $\lambda_2 \equiv \Mp^2 \frac{V''}{V}$ in \eqref{lambda2}, which are plotted in terms of the redshift $z$. They are related to the swampland criterion 2 in \eqref{SC2}. \label{Fig:lambda}} 
\end{figure}

With the effective potentials, now we can check on the second Swampland Criterion in \eqref{SC2},
or say, the refined de Sitter conjecture. We define the following parameters
\begin{align}
\lambda_1 & \equiv M_P \frac{{V'} }{V},\qquad~ \label{lambda1}
V'   \equiv\frac{{\d} V(\phi)}{{\d} \phi}  =\frac{\dot{V}(t)}{\dot{\phi}(t)} \,,   \\ % = \frac{{\d} V( t )/{\d} t }{{\d} \phi(t)/ {\d} t} 
\lambda_2 & \equiv  M_P^2  \frac{{V''} }{V},\qquad \label{lambda2}
V''   \equiv \frac{\d}{{\d} \phi}\frac{{\d}  V(\phi)}{ {\d} \phi}  =\frac{1}{\dot{\phi}(t)}\frac{\d }{\d t} \Big[ \frac{\dot{V}(t)}{\dot{\phi}(t)} \Big].
\end{align}
It is straightforward to plot the numerical result of $\lambda_1(z)$ and $\lambda_2(z)$ in Figure \ref{Fig:lambda}.
Thus, we can see that at the present $z=0$, $\lambda_1(0), \lambda_2(0)\sim {\cal{O}}(1)$, which is the minimum value between $z=0$ and $z=1$.  In the future infinity, both of the metric solutions in the sDGP and hEDU models will approach the de Sitter spacetime. We can see that $\lambda_1(z)|_{z\to-1}\to 0$ from Figure \ref{Fig:lambda}.  It is because we only consider the late time universe, and our effective potential only has the minimum in Figure \ref{figphi}. Thus, the first condition in the second swampland criterion in \eqref{SC2} is satisfied at present, but in tension with the model in the future.

It is interesting to notice that $\lambda_2(z)$ is still non-vanishing at the future infinity $\lambda_2(z)|_{z\to-1}\sim \mathcal{O}(1)$,
which can be tested with the potential in either \eqref{potential1} or \eqref{potential2} . Thus, we can see that in both of the sDGP and {hEDU} models, we still have $\lambda_2(z) \geq  c_3\sim\mathcal{O}(1)$. Or say, near the minimum of the effective potential, we have the condition %maximum or  
\begin{align}
\Mp^2  {|\nabla_i \nabla_j V|}  \geq  c_3 {V} ,\qquad c_3 \sim\mathcal{O}(1).
\label{extend}
\end{align}
Thus, we can suggest that if the condition \eqref{extend} can be included in the refined de Sitter conjecture, then some braneworld models \cite{Maartens:2010ar} with an asymptotic dS spacetime at the future infinity might be included.
Similarly, one can see for example, an interesting embedding of the generalized models of the Randall-Sundrum \cite{Randall:1999ee,Randall:1999vf} braneworld scenarios within string theory has been discussed in \cite{Banerjee:2018qey}.

\vspace{10pt}
%\begin{center}{\bf V. Conclusion}\label{SecVI}\end{center}
%\section{Conclusion}\label{SecV}
\section{Conclusion and Discussion}
\label{secConclusion}
%This set-up can also be implemented in a modified DGP braneworld model.  
 
We study a modified gravity model of the late time accelerating universe, especially the behavior of the universe evolution including the dark sector. We treat the whole dark sector as the holographic dark fluid on the FRW hypersurface in a {flat} bulk \cite{Cai:2017asf}. After using the SNIa and $H_0$ data, we fit a new set of the parameters comparing to the LCDM model. The matter component $\Omega_m$ in Table \ref{tab1} is very small and $\Omega_I$ effectively contributes to the dark sector, including apparent dark matter component. The data fitting matches well with the observations and our theoretical assumption.

 This hEDU model: emergent dark universe model from the holographic viewpoint, can be implemented into the improved sDGP braneworld scenario. We also check the recently proposed swampland criteria on the model and the result is interesting. Especially notice that the potential $V(\phi)$ is asymptotically flat in the far future $z\to -1$. The metric solution is asymptotic de Sitter, which seems to end up in the swampland. 
One can also see that the swampland criterion parameters $\lambda_1$ and $\lambda_2$ are of $\mathcal{O}(1)$ at present $z=0$, but $\lambda_1$ approaches zero and $\lambda_2$ approaches a positive constant in the future infinity. 

Despite this result, we do not think this completely means that the hEDU model is in the swampland. One should notice that we do not have an explicit scalar field and potential. The stress-energy tensor of the holographic fluid which the potential can be derived from is effective at the low energy. The sDGP braneworld model shares similar behavior as the hEDU model in the far future, where the universe is asymptotic de-Sitter, although the current universe satisfies the criteria. One can hope to bring both models back to the string landscape by evoking a phase change of the universe, which is beyond the discussion of the effective low energy behavior presented here.

Another comment we would like to add here is the possible extension of the refined de Sitter conjecture as proposed in (\ref{extend}). The refined conjecture in \cite{Ooguri:2018wrx} is motivated by the distance swampland conjecture,  Bousso's covariant entropy bound and phenomenological counter-examples including Higgs vacuum and pion potentials. We discuss the possible condition (\ref{extend}) here for a phenomenological reason, as it can include certain top-down based brane-world scenarios. An alternative refined de Sitter conjecture conjecture can also be found in \cite{Andriot:2018mav}

% and it means it may also end up in the swampland in the future, 

\section*{\small Acknowledgments}
{\small \footnotesize
We thank Y. S. An, C. C. Han, W. Li, S. Mukohyama, S. Pi, M. Sasaki,  T. Tanaka, S. J. Wang,  H. Wei,  J. Q. Xia, Y. Zhang for helpful conversations.
R.\, -G.\, Cai was supported by the National Natural Science Foundation of China (No.11690022, No.11435006, No.11647601,No. 11851302, and No. 11821505), Strategic Priority Research Program of CAS (No.XDB23030100), Key Research Program of Frontier Sciences of CAS;\,
S. Khimphun was supported by the research fund HEIP(6221-KH), under this Graduate School of Science in Royal University of Phnom Penh;\,
B. -H. Lee was supported by the Basic Science Research Program (No.NRF2018R1D1A1B07048657) through the National Research Foundation(NRF) of Korea funded by the Ministry of education;\,
S.\, Sun was supported by MOST and NCTS in Taiwan, MIUR in Italy under Contract(No. PRIN 2015P5SBHT) and ERC Ideas Advanced
Grant (No. 267985) \textquotedblleft DaMeSyFla";\, G. Tumurtushaa was supported by the Institute for Basic Science (IBS) under the project code(IBS-R018-D1);\,
Y.\, -L.\, Zhang was supported by Grant-in-Aid for JSPS international research fellow(18F18315).
}

\end{document}